\documentclass[12pt]{article}
\usepackage{amssymb,amsmath,epsfig}
\makeatletter

\@addtoreset{equation}{section}
\def\section{\@startsection {section}{1}{\z@}{-2.5ex plus -1ex minus
 -.2ex}{1.3ex plus .2ex}{\large\bf}}
\def\subsection{\@startsection{subsection}{2}{\z@}{-2.25ex plus%
 -1ex minus -.2ex}{0.5ex plus .2ex}{\bf}}

\advance \voffset by -0.8in
\advance \hoffset by -0.6in
\newcommand{\inv}[0]{{-1}}

\newcommand{\hyp}[0]{{\mathbb{H}^2}}

\def\bv{{\mbox{\boldmath $v$}}}

\def\ba{{\mbox{\boldmath $a$}}}

\def\bx{{\mbox{\boldmath $x$}}}

\def\by{{\mbox{\boldmath $y$}}}

\def\bp{{\mbox{\boldmath $p$}}}

\newcommand{\RR}{\mathbb{R}}

\newtheorem{theorem}{Theorem}[section]

\newtheorem{lemma}[theorem]{Lemma}

\newtheorem{definition}[theorem]{Definition}

\def\bea{\begin{eqnarray}}
\def\eea{\end{eqnarray}}
\def\bmz{\left(\begin{array}{2,2}}
\def\emz{\end{array}\right)}
\def\bmd{\left(\begin{array}{3,3}}
\def\emd{\end{array}\right)}

\def\bpm{\begin{pmatrix}}
\def\epm{\end{pmatrix}}

\begin{document}
\parskip 6pt
\parindent 0pt
%\begin{flushright}
%HWM-03-2\\
%EMPG-03-02\\
%gr-qc/yymmnnn
%\end{flushright}

\begin{center}
\baselineskip 24 pt
{\Large \bf   Global Lorentzian geometry from lightlike geodesics: \\
What does an observer in (2+1)-gravity see?}

\baselineskip 18pt

\vspace{1cm}
{\large C.~Meusburger\footnote{\tt  catherine.meusburger@uni-hamburg.de}

Department Mathematik\\
Universit\"at Hamburg\\
Bundestra\ss e 55, D-20146 Hamburg, Germany} \\

\vspace{0.5cm}

{12 January 2010}

\end{center}

\begin{abstract}

\noindent
We show how an observer could measure the non-local holonomy variables  that parametrise the flat Lorentzian 3d manifolds arising as spacetimes in (2+1)-gravity.
 We consider an observer who emits lightrays that return to him at a later time and performs several realistic measurements associated with such returning lightrays: the eigentime elapsed between the emission of the lightrays and their return, the directions into which the light is emitted and from which it returns and the frequency shift between the emitted and returning lightray. 
We show how  the holonomy variables and 
  hence  the full geometry of these manifolds can be reconstructed from these measurements in finite eigentime. 

\end{abstract}

\section{Introduction}
\label{intro}

Gravity in (2+1) dimensions is of interest intrinsically due to its rich mathematical structure and as a toy model for quantum gravity in higher dimensions \cite{Carlipbook}.
Important progress in the understanding of the classical theory and its quantisation followed the discovery  that the theory can be formulated as a Chern-Simons gauge theory \cite{AT,Witten1}. This made it possible to apply gauge theoretical  methods to  the theory and related its quantisation to  the theory of quantum groups,  knot and link invariants and topological quantum field theory \cite{knots,RT}.  

However, it is not straightforward to interpret the resulting quantum theories as quantum geometry or quantised general relativity.  Although the manifolds arising as spacetimes in (2+1)-dimensional gravity have a rich geometrical features  involving Teichm\"uller theory and hyperbolic geometry \cite{mess, npm}, it has been difficult to 
 interpret the associated quantum theories in geometrical terms. This hindered the physical interpretation of the theory and made it difficult to extract interesting physics from (2+1)-gravity. 
 In particular, it remained unclear 
  how the fundamental variables  that parametrise the classical solutions and play a central role in the  quantisation of the theory are related to concrete observations by an observer in the spacetime.

In this paper, we address this issue for classical (2+1)-gravity with vanishing cosmological constant and without matter. The relevant spacetimes are flat, maximally globally hyperbolic  three-dimensional Lorentzian manifolds with a Cauchy surface of genus $g\geq 2$. We consider an observer who investigates the geometry of these spacetimes by emitting lightrays that return to him at a later time. Te observer can perform several measurements associated with such returning lightrays: He can determine the return time, the directions of  into which the lightrays are emitted or from which they return and the relative frequencies of the lightray at its emission and return. 
We give explicit expressions for these measurements in terms of the holonomy variables which play a central role in the quantisation of the theory. Moreover, we demonstrate how these measurements allow the observer to determine the full geometry of the spacetime in finite eigentime.

The paper is structured as follows: Sect.~\ref{backgr}  summarises the relevant properties of 
flat Lorentzian 3d manifolds, their classification and their parametrisation by holonomy variables. In Sect.~\ref{lightmeas} we investigate the measurements associated with returning lightrays. We show how the relevant concepts such as observers, lightrays and returning lightrays are realised in the universal cover. We then derive explicit expressions for  these measurements in terms of the holonomy variables which parametrise the spacetime and discuss their geometrical interpretation. In Sect.~\ref{sptimegeom}, we show how these measurements can be used to  reconstruct the holonomy variables and thus the full geometry of the spacetime from these measurements in finite eigentime. Sect.~\ref{outlook} contains our concluding remarks.

\section{Vacuum spacetimes in (2+1)-gravity}
\label{backgr}

\subsection{Notations and conventions}

We denote by $\mathbb M^3$ three-dimensional Minkowski space with the Minkowski metric  $\eta=\text{diag}(-1,1,1)$. Throughout the paper, we write $\bx\cdot\by$ for $\eta(\bx,\by)$ and $\bx^2$ for $\eta(\bx,\bx)$.  Vectors $\bx\in\RR^3$ are thus timelike, lightlike and spacelike if $\bx^2<0$, $\bx^2=0$ and $\bx^2>0$, respectively. 
The  group  of orientation and time orientation preserving isometries of $\mathbb M^3$ is the three-dimensional Poincar\'e group $P_3=ISO^+(2,1)=SO^+(2,1)\ltimes\RR^3$, which is the semidirect product of the three-dimensional proper, orthochronous Lorentz group $SO^+(2,1)\cong PSL(2,\RR)\cong PSU(1,1)$ with the abelian translation group $\RR^3$. We parametrise elements of $P_3$ as
$(v,\ba)$ with $v\in SO^+(2,1)$ and $\ba\in\RR^3$. In terms of this parametrisation, the group multiplication law then takes the form
\begin{align}
(v_1,\ba_1)\cdot (v_2,\ba_2)=(v_1v_2, \ba_1+v_1\ba_2),
\end{align}
where  $v\ba$ denotes the action of an element $v\in SO^+(2,1)$ on a vector  $\ba\in\RR^3$.
Elements of $SO^+(2,1)$ are called parabolic, elliptic, hyperbolic, respectively, if the stabilise the orthogonal complement  $\by^\bot$ of  a timelike, lightlike or spacelike vector $\by\in\RR^3$.

\subsection{Classification and properties of  (2+1)-spacetimes}

As the Ricci tensor of a three-dimensional  manifold determines its sectional curvature, 
gravity in three dimensions has no local gravitational degrees of freedom. Any vacuum solution
of the three-dimensional Einstein equations without cosmological constant  is a flat Lorentzian 3d manifold which is locally isometric to  Minkowski space $\mathbb M^3$. 
In contrast to the four-dimensional case, this allows one to give an explicit classification of the diffeomorphism equivalence classes of solutions of Einstein's equations. 

A complete classification has been achieved
for  {\em maximally globally hyperbolic} (MGH)  flat Lorentzian 3d manifolds with complete Cauchy surfaces \cite{mess,npm}. The assumption of global hyperbolicity selects spacetimes with acceptable causality behaviour (no closed timelike curves, the intersection of the future of a point with the past of another is always compact).  The condition is equivalent to imposing that the manifold contains a Cauchy-surface, a spacelike two-surface $S$, which every inextensible, causal curve intersects exactly once. The completeness condition  excludes Cauchy surfaces with singularities, while the maximality condition is a technical condition imposed to avoid overcounting of spacetimes.

Throughout the paper, we restrict attention to the simplest case, namely  to MGH flat 3d Lorentzian manifolds $M$ with  a {\em compact} Cauchy surface  $S$ of genus $g\geq 2$\footnote{The case of Cauchy surfaces of genus one (torus universe) can be treated along similar lines, see \cite{Carlipbook, gua} but the geometrical properties of these spacetimes differ from the higher genus cases.}. 
The properties of these manifolds have been determined in \cite{mess}, for a detailed and accessible review see also \cite{bb}. It is shown there that they have  topology
 $M\approx \RR^+\times S$, that they are future complete, but not past-complete and that they have an initial ``big bang" singularity. Moreover, they  are equipped
  with a canonical cosmological time function $T: M\rightarrow \RR^+$ \cite{bg} that tends to zero along every past-directed inextensible causal curve and is given by
 \begin{align}
 T(p)=\text{sup}\{L(c)\;|\; c:[a,b]\rightarrow M, c(a)=p,\; c\; \text{past-directed and causal}\}.
 \end{align}
The surfaces $M_T$ of constant cosmological time are Cauchy surfaces and foliate $M$ 
\begin{align}\label{mfoliate}
M=\bigcup_{T\in\RR^+} M_T.
\end{align}
The classification of flat MGH (2+1)-spacetimes makes use of their description 
 as quotients of their universal covers. It is shown in \cite{mess,barbot}, that their universal covers $\tilde M$ can be identified with regular domains in Minkowski space $\mathbb M^3$. These are open, future complete regions $\tilde M\subset\mathbb M^3$ that are domains of dependence, i.~e.~given as the future of a set $\tilde M_0\subset \mathbb M^3$. The cosmological time function on $M$ lifts to a cosmological time function $T:\tilde M\rightarrow \RR^+$, which gives the geodesic distance of points $p\in\tilde M$ from $M_0$ and whose level surfaces foliate the domain
\begin{align}
\tilde M=\bigcup_{T\in\RR^+} \tilde M_T.
\end{align}
The fundamental group $\pi_1(M)\cong\pi_1(S)$ acts on $\tilde M$ via deck transformations,  which are given by a group homomorphism $h: \pi_1(M)\rightarrow P_3$, in the following referred to as holonomies. This group action is free and properly discontinuous and preserves each surface $\tilde M_T$ of constant cosmological
time. Moreover, the Lorentzian component of the holonomies $h_L: \pi_1(M)\rightarrow SO^+(2,1)$
defines a faithful and discrete representation of $\pi_1(M)$. For the case of a Cauchy surface of genus $g\geq 2$, this implies that the image of $h_L$ is a cocompact Fuchsian group $\Gamma\cong\pi_1(M)\cong\pi_1(S)$ of genus $g\geq 2$. This is a discrete subgroup of the three-dimensional Lorentz group $SO^+(2,1)\cong PSL(2,\RR)$ with $2g$ generators and a defining relation
\begin{align}
\Gamma=\langle v_{a_1},v_{b_1},...,v_{a_g},v_{b_g}\;|\; [v_{b_g}, v_{a_g}]\cdots[v_{a_1}, v_{b_1}]=1\rangle\subset PSL(2,\RR).
\end{align}
Note that all elements of such a cocompact Fuchsian group are hyperbolic, i.~e.~they stabilise planes in $\mathbb M^3$ with {\em spacelike} normal vectors.

It has been shown by Mess \cite{mess} that the holonomies characterise flat MGH Lorentzian 3d manifolds  with compact, complete Cauchy surfaces uniquely. Given a group homomorphism $h:\pi_1(M)\rightarrow P_3$ 
whose Lorentzian component $h_L: \pi_1(M)\rightarrow SO^+(2,1)$ defines a cocompact Fuchsian group of genus $g$, there exists a unique  domain $\tilde M\subset \mathbb M^3$ on which $\pi_1(M)\cong\pi_1(S)$ acts freely and properly discontinuously in such a way that each constant cosmological time surface is preserved. 

Two flat MGH spacetimes with Cauchy surfaces of genus $g\geq 2$ are isometric if and only if the associated holonomy maps are related by conjugation with a constant element of the Poincar\'e group $P_3$.
This implies that the physical (or reduced) phase space of the theory, the set of diffeomorphism equivalence classes of solutions of the three-dimensional Einstein equations, can be identified with
\begin{align}\label{phsp}
\mathcal P=\text{Hom}_0(\pi_1(M), P_3)/P_3,
\end{align}
where the index 0 indicates that the Lorentzian component of the group homomorphism must  define a Fuchsian group of the appropriate genus. Note that $\mathcal P$ is a connected component of the moduli space of flat  $P_3$-connections on $S$ and coincides with the cotangent bundle
$T^*\tau(S)$ of Teichm\"uller space on $S$. The holonomy variables $h(\lambda)$, $\lambda\in\pi_1(M)$, thus play a central role in the description of the solutions of the theory and in its quantisation.

\subsection{Geometry of (2+1)-spacetimes}

It is shown in \cite{mess,bb} that the description of these manifolds as quotients of their universal covers gives rise to a concrete and explicit description of their geometry. One distinguishes two cases, conformally static spacetimes, for which the translational components of the holonomies are trivial, and evolving ones for which this is not the case.

\subsubsection{Conformally static spacetimes}

For conformally static spacetimes $M^s$, the universal cover $\tilde M^s$ is the interior of the future lightcone of a point $\bp\in\mathbb M^3$
\begin{align}\label{statdom}
\tilde M^s=\{\by\in\mathbb  M^3\;|\; (\by-\bp)^2<0, y^0-p^0>0\}.
\end{align}
The cosmological time function gives the Lorentzian distance from  the tip of the lightcone
\begin{align}\label{statct}
T(\by)=\sqrt{|(\by-\bp)^2|}\qquad\forall \by\in\tilde M^s,
\end{align}
and the foliation of $\tilde M^s$ by constant cosmological time surfaces is the standard foliation of the lightcone by hyperboloids. Each constant cosmological time surface $\tilde M_T^s$ is thus a copy of two-dimensional hyperbolic space $\hyp$, rescaled with the cosmological time $T$
\begin{align}\label{statsurf}
\tilde M_T^s=\{\by\in\mathbb M^3\;|\; (\by-\bp)^2=-T^2,\;y^0-p^0>0\}\cong T\cdot \hyp.
\end{align}
 As the surfaces $\tilde M_T^s$ must be preserved by the action of $\pi_1(M^s)$ via the holonomies $h: \pi_1(M^s)\rightarrow P_3$, the translational component of the holonomies is trivial. It is given by  conjugation with a global translation  to the tip of the lightcone 
\begin{align}\label{conjhol}
h(\lambda)=(1,\bp)\cdot (v_\lambda,0)\cdot (1,-\bp)\qquad\forall \lambda\in\pi_1(M^s),
\end{align}
while the Lorentzian components $v_\lambda$, $\lambda\in\pi_1(M^s)$, define a cocompact Fuchsian group $\Gamma$ of genus $g$. The action of  $\pi_1(M^s)$ on the constant cosmological time surfaces $\tilde M_T^s$ agrees with the canonical action of $\Gamma$ on $\hyp$.
Each constant cosmological time surface $M_T^s$ in the quotient spacetime 
is thus a copy of  the same Riemann surface $\Sigma_\Gamma=\hyp/\Gamma$,  rescaled with the cosmological time and equipped with a metric of constant curvature $-1/T^2$
\begin{align}\label{confst}
M^s=\bigcup_{T\in\RR^+}  M^s_T=\bigcup_{T\in\RR^+} T\cdot \Sigma_\Gamma \qquad\qquad g_s=-dT^2+T^2g_{\Sigma_\Gamma}.
\end{align}
As the geometry of the constant cosmological time surfaces $M_T$ does not evolve with$T$ and the metric takes the form \eqref{confst}, these spacetimes are called conformally static.

\subsubsection{Evolving spacetimes via grafting}

It has been shown by Mess \cite{mess} that any  evolving flat MGH 3d Lorentzian manifold with a  Cauchy surface of genus $g\geq 2$ can be obtained from a conformally static one via grafting along a measured geodesic lamination on the associated Riemann surface $\Sigma_\Gamma$. 
We sketch the grafting construction for 
 measured geodesic laminations which are weighted multicurves, i.~e.~sets of non-intersecting, simple geodesics on $\Sigma_\Gamma$, each associated with a positive number, the weight. General measured geodesic laminations on a Riemann surface $\Sigma_\Gamma$  are obtained as limits of suitable sequences of weighted multicurves \cite{mess,bb}.

\begin{figure}[h]
\centering
\includegraphics[scale=0.3]{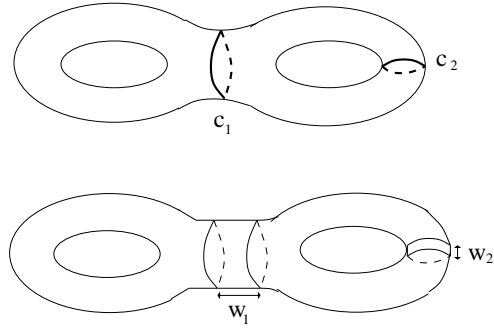}
\caption{\small{Grafting along  two geodesics on a genus two surface.}}
\label{graftsurf}
\end{figure}

Schematically, grafting on a Riemann surface $\Sigma_\Gamma$ amounts to cutting $\Sigma_\Gamma$ along all geodesics in the weighted multicurve and gluing in strips 
whose width is given by the weight, as shown in Figure \ref{graftsurf}. While $\Sigma_\Gamma$ is equipped with a Riemannian metric of constant curvature -1, the metric on the grafted strips is a Riemannian metric of vanishing curvature.

\begin{figure}[h]
\centering
a)\includegraphics[scale=0.4]{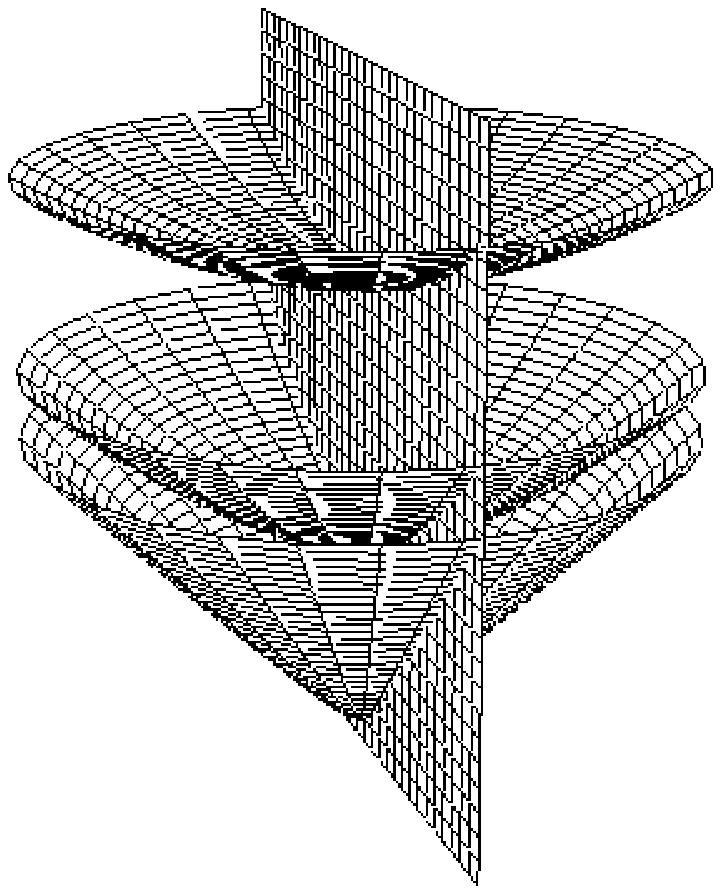}
b)\includegraphics[scale=0.4]{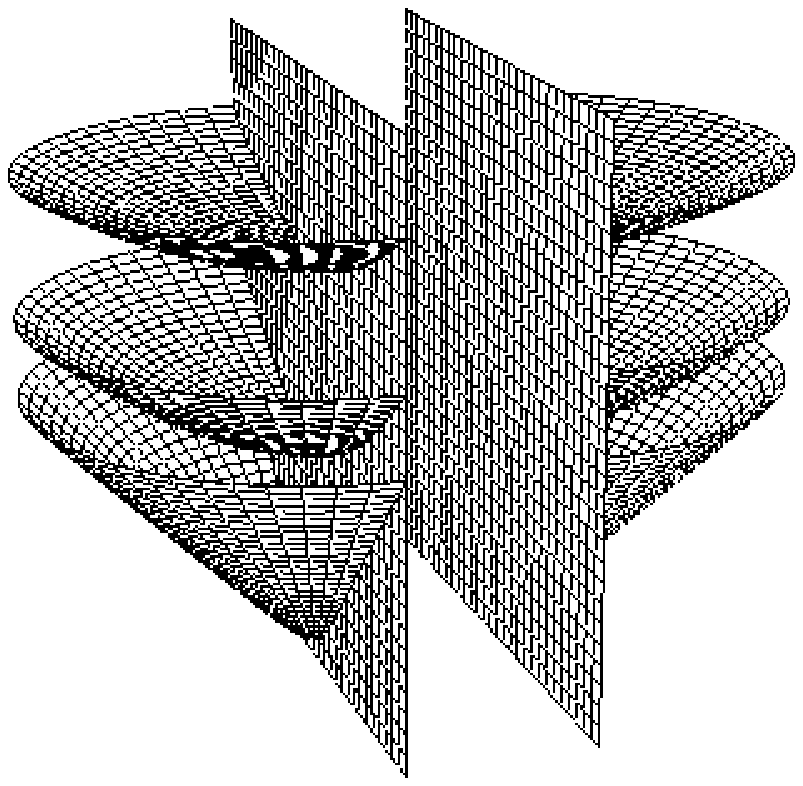}
c)\includegraphics[scale=0.4]{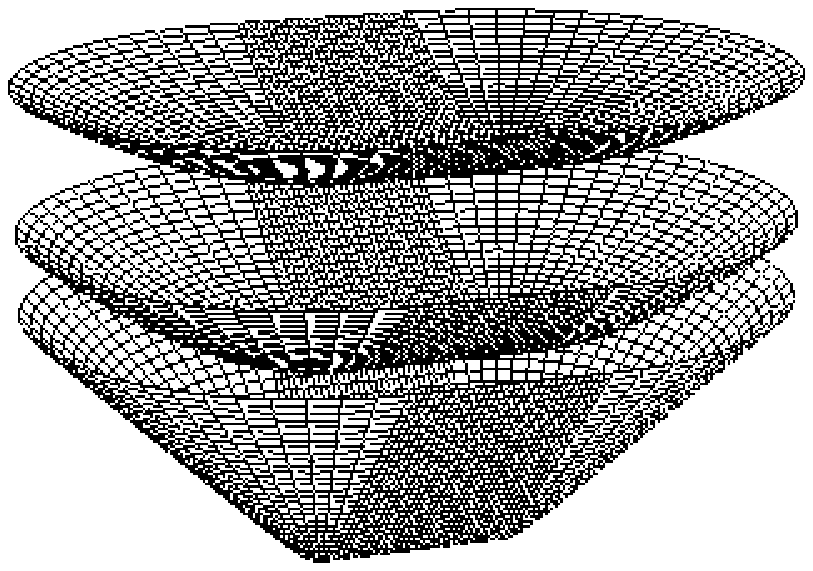}
\caption{\small{The grafting construction  in the lightcone for a single geodesic.}}
\label{grafting}
\end{figure}

Grafted  (2+1)-spacetimes are obtained  by applying the grafting construction simultaneously to all constant cosmological time surfaces $M_T^s=T\cdot\Sigma_\Gamma$ such that the weight of the grafting geodesics  is the same for all surfaces $M_T^s$. 
The construction is performed in the universal cover. For this, one lifts all grafting geodesics on $\Sigma_\Gamma$ to a weighted multicurve on each constant cosmological time surface $\tilde M_T^s\cong T\cdot\hyp$ by taking one lift of each geodesic on $\Sigma_\Gamma$ and acting on it with the holonomies. This yields an infinite set of non-intersecting weighted geodesics on each hyperboloid $\tilde M_T^s$. These geodesics are given as the intersection of the lightcone $\tilde M^s$ with planes with spacelike normal vectors as shown in Figure \ref{grafting} a). 

To construct the grafted domain, one
 selects a basepoint  in $\tilde M^s$ outside all geodesics in the  lifted multicurves. One then cuts the lightcone  along all planes defined by these geodesics and shifts the pieces that do not contain the basepoint away from the basepoint, in the direction of the planes' normal vectors and by a distance given by the weights as shown in Figure \ref{grafting} b). Finally, one joins the translated pieces by straight lines as indicated in Figure \ref{grafting} c).

This yields a deformed domain, which is no longer the future of a point but of a graph $\tilde M_0$. The cosmological time function of the deformed domain gives the geodesic distance of points in $\tilde M$ from $\tilde M_0$. The surfaces $\tilde M_T$ of constant cosmological time are the images of the hyperboloids $\tilde M_T^s$ under the grafting construction. They are  deformed hyperboloids with strips glued in along the geodesics in the multicurve. 

The fundamental group $\pi_1(M)$ acts on the deformed domain $\tilde M$ in such a way that two points  on a constant cosmological time surface are identified if and only if the corresponding points on the hyperboloid are identified for the associated conformally static spacetime. The holonomies $h: \pi_1(M)\rightarrow P_3$ thus acquire a non-trivial translational component which takes into account the translations  in the grafting construction. 

\begin{figure}[h]
\centering
\includegraphics[scale=0.4]{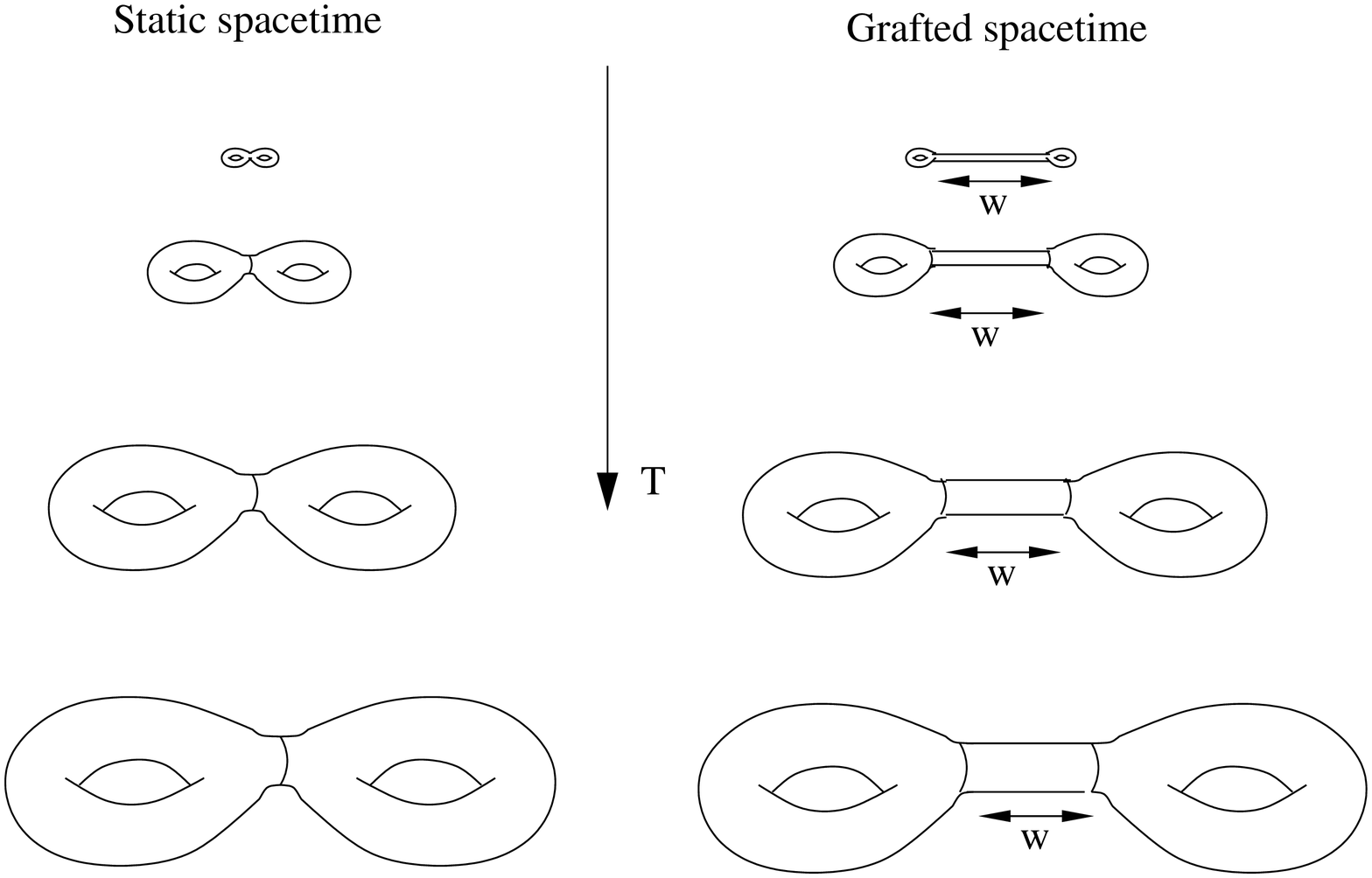}
\caption{\small{Evolution of static and grafted spacetimes with the cosmological time.  The hyperbolic part of each constant cosmological time surface  is rescaled with the cosmological time $T$, the width of the grafted strips remains constant.}}
\label{timevol}
\end{figure}

The spacetime $M$ is obtained by taking the quotient of the deformed domain $\tilde M$ by this action of the fundamental group. One finds that its constant cosmological time surfaces $M_T$  undergo a non-trivial evolution with the cosmological time $T$, which is indicated in Figure \ref{timevol}. While the hyperbolic part of the constant cosmological time surfaces, i.~e.~the part outside of the grafting strips, is rescaled with the cosmological time $T$, the width of the grafted strips  remains constant. The effect of grafting is thus dominant  near the initial singularity for $T\rightarrow 0$ and vanishes in the limit $T\rightarrow\infty$.

\section{Measurements associated with lightrays}
\label{lightmeas}

\subsection{Spacetime geometry via returning lightrays}
\label{idea}

While the description of  flat 3d MGH Lorentzian  spacetimes as quotients of their universal covers allows one to classify these spacetimes and to give an explicit description of their geometry, it does not provide a clear physical interpretation. To extract interesting physics from the theory, one needs to relate the  variables which characterise these manifolds  to concrete geometrical  quantities that could be measured by an observer in the spacetime. 
These measurements should allow the observer to distinguish different spacetimes and to determine their geometry in finite eigentime.
 
This is difficult, since measurements are required to be {\em local}, i.~e.~take place in a small neighbourhood of a point in the spacetime, while the manifolds under consideration are locally isometric to Minkowski space $\mathbb M^3$ and can be distinguished only through their {\em global} geometrical properties. Moreover, as the spacetimes contain no matter and their curvature vanishes everywhere,  it is a priori not clear what physically meaningful measurements could be performed at all.

The  idea that allows one to address this problem 
 is to consider an observer who probes the geometry of the spacetime by emitting lightrays. As we will see in the following, lightrays emitted in certain directions  return to the observer at a later time. The observer can then perform several measurements associated with such returning lightrays. He can record their return time, the amount of eigentime elapsed between the emission of the lightray and its return. He can determine the directions into which light is emitted and from which it returns, 
 and he can compare the frequencies of the emitted and returning  lightrays.

\subsection{Physics with returning lightrays}
\label{retlray}

To derive explicit expressions for the measurements associated with returning lightrays in a flat MGH 3d Lorentzian manifold $M$, it is advantageous to work in its universal cover, the associated  domain $\tilde M\subset\mathbb M^3$. This requires a clear definition
of the relevant concepts such as observers, lightrays and returning lightrays in terms of the universal cover.  We start with the notion of an observer, restricting attention to observers in free fall.

\begin{definition} (Observer) \label{obs}

An  {\em  observer} (in free fall) in $M$ is characterised uniquely by a timelike, future-directed geodesic $g: [a,\infty)\rightarrow M$, his {\em worldline}. Equivalently, an observer can be defined as a $\pi_1(M)$-equivalence class   of timelike, future-directed geodesics $\tilde g: [a,\infty)\rightarrow \tilde M$ in the universal cover with the equivalence relation  $\tilde g_1\sim_{h} \tilde g_2$ if there exists an element $\lambda\in\pi_1(M)$ such that $\tilde g_1(t)=h(\lambda)\tilde g_2(t)$ for all $t\in[a,\infty)$. 
The worldline $g:[a,\infty)\rightarrow M$ is  {\em parametrised according to eigentime} if $\dot g(t)^2=-1$ $\forall t\in[a,\infty)$ or, equivalently, 
 $ \dot{\tilde g}(t)^2=-1$ $\forall t\in[a,\infty)$ for all lifts $\tilde g$ of $g$.
\end{definition}
If  an observer's worldline $g:t\mapsto g(t)$ is parametrised according to eigentime, the parameter $t$ gives coincides with the  time as perceived by the observer, i.~e.~the time that would be shown by a clock carried with him.  Note that the eigentime is unique up to a constant time shift $t\mapsto t+t_0$ with  $t_0\in\RR$. 

Analogously, we can define a lightray as a $\pi_1(M)$-equivalence class of geodesics in the universal cover. This yields the following definition.
\begin{definition}\label{lray} (Lightray)

A {\em lightray} in $M$ is a lightlike, future-directed geodesic $c: [p,\infty)\rightarrow M$ or, equivalently, a $\pi_1(M)$-equivalence class of lightlike, future-directed geodesics in $\tilde M$ with  equivalence relation $\sim_h$ of Def.~\ref{obs}. A lightray emitted (received) by an observer with worldline  $g:[a,\infty)\rightarrow M$ at eigentime $t$ is a lightlike, future-directed geodesic $c:[p,q]\rightarrow M$ with $c(p)=g(t)$ $(c(q)=g(t))$ or, equivalently, the $\pi_1(M)$-equivalence class of lightlike, future-directed geodesics $\tilde c:[p,q]\rightarrow\tilde M$ for which there exists a lift $\tilde g:[a,\infty)\rightarrow \tilde M$ of  $g$ such that $\tilde c(p)=\tilde g(t)$ $(\tilde c(q)=\tilde g(t))$.
\end{definition}

Note that the picture obtained by lifting timelike or lightlike curves to the universal cover differs from the one habitually encountered in Riemannian geometry, where a {\em closed} curve on a Riemannian manifold lifts to a single {\em open} curve in its universal cover. As the spacetimes under consideration are globally hyperbolic, they do not exhibit any closed time-  or lightlike curves. Instead, open time- or lightlike curves in $M$  lift to a $\pi_1(M)$-equivalence class of time- or lightlike curves in $\tilde M$. This accounts for the possibility of {\em returning lightrays}, lightrays emitted by an observer  that return to him at a later time. 

\begin{definition}\label{retlray} (Returning lightray)

A {\em returning lightray} with respect to an observer  with worldline $g:[a,\infty)\rightarrow M$ is a lightlike, future-directed geodesic $c:[p,q]\rightarrow M$ that intersects $g$ in $c(p)$ and $c(q)$. Equivalently, a returning lightray is given as a $\pi_1(M)$-equivalence class of lightlike, future-directed geodesics $\tilde c:[p,q]\rightarrow \tilde M$ such that there exists an element $\lambda\in\pi_1(M)$ and a lift  $\tilde g:[a,\infty)\rightarrow \tilde M$  of $g$ with $\tilde c(p)\in \tilde g$ , $\tilde c(q)\in h(\lambda)\tilde g$.\end{definition}

\begin{figure}[h]
\centering
\includegraphics[scale=0.35]{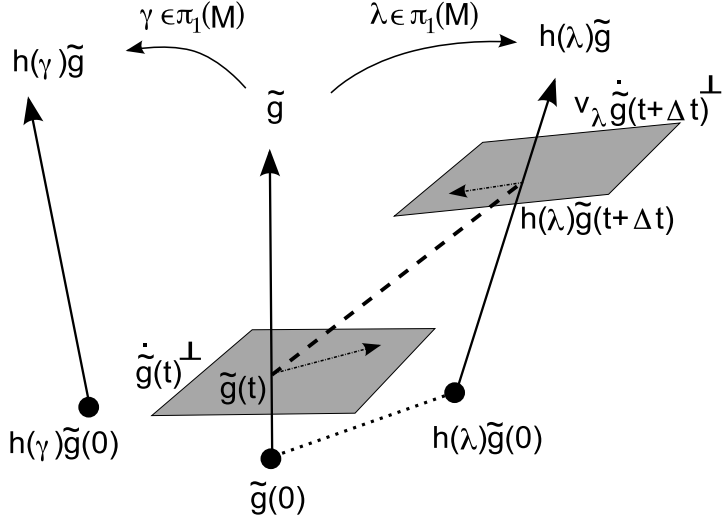}
\caption{\small{Lifts of the observer's worldline to $\tilde M$ with returning lightray (dashed line), orthogonal complements $\dot{\tilde g}^\bot(t)$, $(v_\lambda\dot{\tilde g})(t+\Delta t)^\bot$ (grey planes) and projection of the returning lightray to  $\dot{\tilde g}^\bot(t)$, $(v_\lambda\dot{\tilde g})(t+\Delta t)^\bot$  (dashed arrows).}}
\label{lrdomain}
\end{figure}

The description of returning lightrays in the universal cover $\tilde M$ is pictured in Figure \ref{lrdomain}.  As a returning lightray relates a lift of the observer's worldline to one of its images under the action of $\pi_1(M)$, it defines a unique element $\lambda\in\pi_1(M)$. However, it is a priori not guaranteed that  for each observer and each element of $\pi_1(M)$ there exists an associated returning lightray. This is a consequence of the geometry of Minkowski space and the fact that the domains $\tilde M$ are future-complete.

\begin{lemma} \label{retlrexist}
Let $g:[a,\infty)\rightarrow M$ be the worldline of an observer in free fall. Then for all $t\in[a,\infty)$ the returning lightrays $c:[p,q]\rightarrow M$ with $c(p)=g(t)$ are in one-to-one correspondence with elements of the fundamental group $\pi_1(M)$. 
\end{lemma}

{\bf Proof:}  We consider a  lift $\tilde g:[a,\infty)\rightarrow \tilde M$ of the observer's worldline. Returning lightrays $c:[p,q]\rightarrow M$ with $c(p)=g(t)$ are in  one-to-one  correspondence with lightlike geodesics $\tilde c_\lambda:[p,q]\rightarrow \tilde M$ such that $\tilde c_\lambda(p)=\tilde g(t)$ and $\tilde c_\lambda(q)\in h(\lambda)\tilde g$  for a certain $\lambda\in\pi_1(M)$. As $\tilde g$ and $h(\lambda)\tilde g$ are timelike, future oriented geodesics in Minkowski space, for all choices of 
$\lambda\in\pi_1(M)$, $t\in[a,\infty)$,
there exists a a lightlike geodesic $\tilde c_\lambda:[p,q]\rightarrow \mathbb M^3$ such that $\tilde c_\lambda(p)=\tilde g(t)$ and $\tilde c_\lambda(q)\in h(\lambda)\tilde g$. As $\tilde g(t)\in\tilde M\subset\mathbb M^3$ and $\tilde M$ is future complete, we have $\tilde c_\lambda(t)\in\tilde M$ for all $t\in[p,q]$, and $\tilde c_\lambda$ defines a returning lightray. 
\hfill $\Box$

After formulating the relevant concepts  in terms of the universal cover $\tilde M\subset \mathbb M^3$, we are ready to consider the measurements associated with returning lightrays.  We start with the return time, the interval of eigentime elapsed between the emission of a returning lightrays and its return as measured by the observer.

\begin{definition}(Return time) \label{rettime}

Let $g:[a,\infty)\rightarrow M$ be the worldline of an observer  in free fall, parametrised according to eigentime and $\tilde g:[a,\infty)\rightarrow \tilde M$ a lift of $g$. 
Then by Lemma \ref{retlrexist} for each $t_e\in[a,\infty)$ and each $\lambda\in\pi_1(M)$ there exists a unique $t_r\in(t_e,\infty)$ and a unique lightlike geodesic $\tilde c_\lambda:[0,1]\rightarrow \tilde M$ with $\tilde c_\lambda(0)=\tilde g(t_e)$ and $\tilde c_\lambda(1)=h(\lambda)\tilde g(t_r)$. 
The return time is given by $\Delta t=t_r-t_e$ and obtained as the unique positive solution of the quadratic equation
\begin{align}\label{condi}
(h(\lambda)\tilde g(t_e+\Delta t)-\tilde g(t_e))^2=0.
\end{align}
\end{definition}

To obtain the directions in which the light needs to be emitted in order to return to the observer and from which it returns, we recall that the directions an observer perceives as ``spatial" are given by the orthogonal complement $\dot{g}(t)^\bot$, where $g:[a,\infty)\rightarrow M$ is the observer's wordline. By considering the associated quantities in the universal cover, we obtain
the following definition.

\begin{definition} \label{directions} (Directions of emission and return)

Let $g:[a,\infty)\rightarrow M$ the worldline of an observer, parametrised according to eigentime, and $\tilde g:[a,\infty)\rightarrow\tilde M$ a lift of $g$. Let $\tilde c:[p,q]\rightarrow\tilde M$ be a future-directed, lightlike geodesic with $\tilde g(t)=\tilde c(p)$  $(\tilde g(t)=\tilde c(q))$. Then the direction into which the lightray associated with $\tilde c$ is emitted (from which the lightray  lightray associated with $\tilde c$ returns) as perceived by the observer is given by the spacelike
unit vector
\begin{align}\label{emdir}
\hat \bp_e=\Pi_{\dot {\tilde g}(t)^\bot}(\dot {\tilde c}(p))/{|\Pi_{\dot {\tilde g}(t)^\bot}(\dot {\tilde c}(p))|}\qquad \left(\hat \bp_r=\Pi_{\dot {\tilde g}(t)^\bot}(\dot {\tilde c}(q))/{|\Pi_{\dot {\tilde g}(t)^\bot}(\dot {\tilde c}(q))|}\right).
\end{align}
\end{definition}

Finally, the observer can determine the  relative frequencies of a returning lightray at its emission and return. In the universal cover, this problem is analogous to the relativistic Doppler effect. The only difference is  that here the two timelike geodesics correspond to a lift of the observer's wordline and its image instead of two different observers for the relativistic Doppler effect.   This yields the following definition.

\begin{definition}\label{doppler} (Frequency shift)

Let $g: [a,\infty)\rightarrow M$ be the worldline of an observer parametrised according to eigentime and $\tilde g:[a,\infty)\rightarrow \tilde M$ a lift of $g$. Let $\tilde c:[p,q]\rightarrow \tilde M$ a lightlike geodesic associated with a  returning lightray with $\tilde c(p)=\tilde g(t_e)$, $\tilde c(q)=h(\lambda)\tilde g(t_r)$ for an element $\lambda\in\pi_1(M)$. Then the quotient of  frequencies of the lightray at its emission and return as measured by the observer is given by
\begin{align}\label{frequdef}
\frac{f_r}{f_e}=\frac {h(\lambda)\dot {\tilde g}\cdot (h(\lambda)\tilde g(t_r)-\tilde g(t_e))}{\dot {\tilde g}\cdot (h(\lambda)\tilde g(t_r)-\tilde g(t_e))}.
\end{align}
\end{definition}

\subsection{Explicit results}

To obtain explicit results for the return time, the directions of emission and return and the frequency shift, we make use of the fact that the universal cover $\tilde M$ is a future-complete region in Minkowski space $\mathbb M^3$. The geodesics characterising the observer and the lightrays therefore take a particularly simple form. In the following, we parametrise
 timelike, future-directed geodesics in $\tilde M$  in terms of an element $\bx\in\mathbb H^2$, the {\em velocity vector} and a vector $\bx_0\in\tilde M$, its {\em initial position} at $t=0$
\begin{align}
\tilde g(t)=t\bx+\bx_0\qquad \bx^2=-1, x^0>0, \bx_0\in \tilde M\label{obsparam}.
\end{align}
Note that the parametrisation is unique up to a time shift
\begin{align}\label{timeshift}
t\mapsto t-t_0\qquad \bx_0\mapsto \bx_0+t_0\bx.
\end{align}
Similarly, each lightlike, future-directed geodesic $\tilde c: [0,\infty)\rightarrow\tilde M$ is given by a lightlike vector $\by$ and a initial position vector $\by_0\in\tilde M$ 
\begin{align}
\tilde c(s)=s \by+\by_0\qquad \by^2=0, y^0>0,\by_0\in \tilde M\label{lightparam}.
\end{align}
To obtain explicit expressions for the measurements performed by the observer, it is advantageous to introduce additional parameters, which are given as functions of the velocity vector $\bx$, the initial position $\bx_0$ and the holonomies  $h(\lambda)=(v_\lambda, \ba_\lambda)$. For $\lambda\in\pi_1(M)\setminus\{1\}$ and $\tilde g, \bx,\bx_0$ as in \eqref{obsparam}, we define 
\begin{align}\label{rholdef}
&\cosh\rho_\lambda=-\bx\cdot v_\lambda\bx\\
\label{pardef}
&h(\lambda)\tilde g(0)-\tilde g(0)= \sigma_\lambda(v_\lambda\bx-\bx) +\tau_\lambda v_\lambda\bx+\nu_\lambda \bx\wedge v_\lambda\bx.
\end{align}
The parameter $\rho_\lambda$, which depends only on the velocity vector $\bx$ and the Lorentzian component of the holonomy has a direct interpretation as the geodesic distance $\rho_\lambda=d_{\hyp}(\bx, v_\lambda \bx)$ of $\bx$ and $v_\lambda\bx$ in hyperbolic space $\hyp$. It coincides with the length of the associated geodesic on the Riemann surface $\hyp/\Gamma$. 
The parameters $\sigma_\lambda, \tau_\lambda, \nu_\lambda$ characterise the relative initial position of the geodesic $\tilde g$ and its image $h(\lambda)\tilde g$. They
depend on the velocity vector $\bx$ , the initial position $\bx_0$ as well as both components of  the holonomies. They are invariant under Poincar\'e transformations $(v,\ba)$ acting simultaneously on the geodesic $\tilde g$ and on  all holonomies by conjugation
\begin{align}\label{globpoinc}
\bx\mapsto v\bx,\; \bx_0\mapsto v\bx_0+\ba\qquad h(\lambda)\mapsto (v,\ba)\cdot h(\lambda)\cdot (v,\ba)^\inv.
\end{align}
Using the Definitions \ref{rettime}, \ref{directions}, \ref{doppler} one can derive explicit expressions for the measurements associated with returning lightrays in terms of the parameters $\rho_\lambda,\sigma_\lambda, \tau_\lambda, \nu_\lambda$. These expressions are derived in \cite{icht} and summarised in the following theorem.

\begin{theorem} \cite{icht}
\label{concvalues}

Let $\tilde g: [a,\infty)\rightarrow \tilde M$ be a lift of the worldline of an observer in free fall, parametrised as in \eqref{obsparam}. Consider a returning lightray associated with an element $\lambda\in\pi_1(M)$ that is emitted by the observer at eigentime $t$ and returns at $t+\Delta t$ . 
Then the eigentime $\Delta t$ elapsed between the emission and return of the lightray is given by
\begin{align}\label{retform}
\Delta t (t,\bx,\bx_0, h(\lambda))=(t+\sigma_\lambda)(\cosh \rho_\lambda -1)-\tau_\lambda+\sinh\rho_\lambda\sqrt{(t+\sigma_\lambda)^2+\nu_\lambda^2},
\end{align}
where $\rho_\lambda,\sigma_\lambda,\tau_\lambda,\nu_\lambda$ are functions of $\bx,\bx_0$ and $h(\lambda)$ defined by \eqref{rholdef}, \eqref{pardef}. The direction into which the lightray is emitted is given by the spacelike unit vector
\begin{align}\label{dirvec}
&\hat \bp^e_\lambda=\cos\phi_e \;\frac{v_\lambda\bx+(\bx\cdot v_\lambda\bx)\bx}{|v_\lambda\bx+(\bx\cdot v_\lambda\bx)\bx|}+\sin\phi_e\; \frac{\bx\wedge v_\lambda\bx}{|\bx\wedge v_\lambda\bx|}\\
&\label{dir}
\tan \phi_e(t,\bx,\bx_0,h(\lambda))=\frac{\nu_\lambda}{\sinh\rho_\lambda\sqrt{(t+\sigma_\lambda)^2+\nu_\lambda^2}}
\end{align}
and the direction from which it returns by
\begin{align}\label{dirvec2}
&\hat \bp^r_\lambda(t)=\cos\phi_r \; \frac{v_\lambda^\inv \bx+(\bx\cdot v_\lambda^\inv \bx) \bx}{|v_\lambda^\inv \bx+(\bx\cdot v_\lambda^\inv \bx) \bx|}+\sin\phi_r\; \frac{\bx\wedge v_\lambda^\inv\bx}{|\bx\wedge v_\lambda^\inv\bx|}\\
&\tan\phi_r(t,\bx,\bx_0,h(\lambda))=\frac{\nu_\lambda}{(t+\sigma_\lambda)}\label{dir2}.
\end{align}
The relative frequencies of the lightray at its emission and return as measured by the observer are given by
\begin{align}\label{frequshift}
f_r/f_e(t,\bx,\bx_0, h(\lambda))=\frac{\sqrt{(t+\sigma_\lambda)^2+\nu_\lambda^2}}{\cosh\rho_\lambda\sqrt{(t+\sigma_\lambda)^2+\nu_\lambda^2}+\sinh\rho_\lambda(t+\sigma_\lambda)}<1.
\end{align}
\end{theorem}

As indicated by the notation, the return time, directions and frequency shift are given as functions of the emission time $t$, the two vectors $\bx,\bx_0$ characterising the observer's worldline and the holonomies $h(\lambda)$, $\lambda\in\pi_1(M)$.  The fact that they depend only on the sum $t+\sigma_\lambda$, but not on $t$ and $\sigma_\lambda$ individually, reflects the invariance under a time shift \eqref{timeshift}. Moreover, as they are given in terms of the parameters $\rho_\lambda$, $\sigma_\lambda$, $\tau_\lambda$, $\nu_\lambda$ formulas \eqref{retform},\eqref{dir}, \eqref{dir2}  and \eqref{frequshift} are invariant under Poincar\'e transformations \eqref{globpoinc} acting simultaneously on the observer's geodesic and on  the holonomies. In particular, this implies that they are invariant under change of the choice of the lift combined with an inner automorphism of $\pi_1(M)$
\begin{align}\label{changeoflift}
\tilde g\mapsto h(\eta)\tilde g\qquad \lambda\mapsto \eta\cdot\lambda\cdot \eta^\inv \quad \forall \lambda\in\pi_1(M).
\end{align}

\subsection{Interpretation}

To understand how the expressions \eqref{retform} to \eqref{frequshift} reflect the geometry of the underlying  spacetime, we consider a conformally static spacetime and an observer whose worldline starts at $t=0$ at the initial singularity. In this case, the universal cover can be identified with the lightcone based at the origin, and the translational components of the holonomies as well as  the initial position vector $\bx_0$ in \eqref{obsparam} can be set to zero. This implies that the parameters $\sigma_\lambda, \tau_\lambda, \nu_\lambda$ in \eqref{pardef} vanish for all $\lambda\in\pi_1(M)$ and the expressions for the return time, directions and the frequency shift take the form
\begin{align}\label{statvalue}
&\Delta t(t,\bx,\bv_\lambda)=t(e^{\rho_\lambda}-1)\qquad \phi^{e,r}(t,\bx,\bv_\lambda)=0\qquad
f_r/f_e(t,\bx,\bv_\lambda)=e^{-\rho_\lambda}.
\end{align}
The return time is thus a linear function of eigentime at which the lightrays was emitted, with a coefficient related to the length $\rho_\lambda$ of the associated geodesic on a constant cosmological time surface. The directions of emission and return as well as the frequency shift are independent of the emission time. Note that the frequency shift is a red shift $f_r/f_e<1$, as  expected for an expanding spacetime,  and depends only on the length $\rho_\lambda$ of the associated geodesic on a constant cosmological time surface. 

For a general spacetimes and  general observers, these values of the return time, directions and frequency shift are approached in the limit $t\rightarrow\infty$
\begin{align}
\lim_{t\rightarrow\infty} \Delta t/t=e^{\rho_\lambda}-1\qquad \lim_{t\rightarrow \infty} \phi^{e,r}=0\qquad\lim_{t\rightarrow \infty} f_r/f_e=e^{-\rho_\lambda}
\end{align}
This reflects the fact that for all observers, the cosmological time tends to infinity $T(\tilde g(t))\rightarrow \infty$ as $t\rightarrow\infty$. In this limit, the 
 effects of grafting  become negligible and the spacetime approaches the associated conformally static spacetime. 

Generally, for a given element  $\lambda\in\pi_1(M)$, the return time \eqref{retform} is a linear function of the emission time and the directions of emission \eqref{dirvec} and return \eqref{dirvec2} as well as the frequency shift  \eqref{frequshift}  are constant  if and only if the parameter $\nu_\lambda$  defined by  \eqref{pardef} vanishes.  It is shown in \cite{icht} that this reflects the geometrical properties of the grafting construction.

\begin{figure}[h]
\centering
a)\includegraphics[scale=0.3]{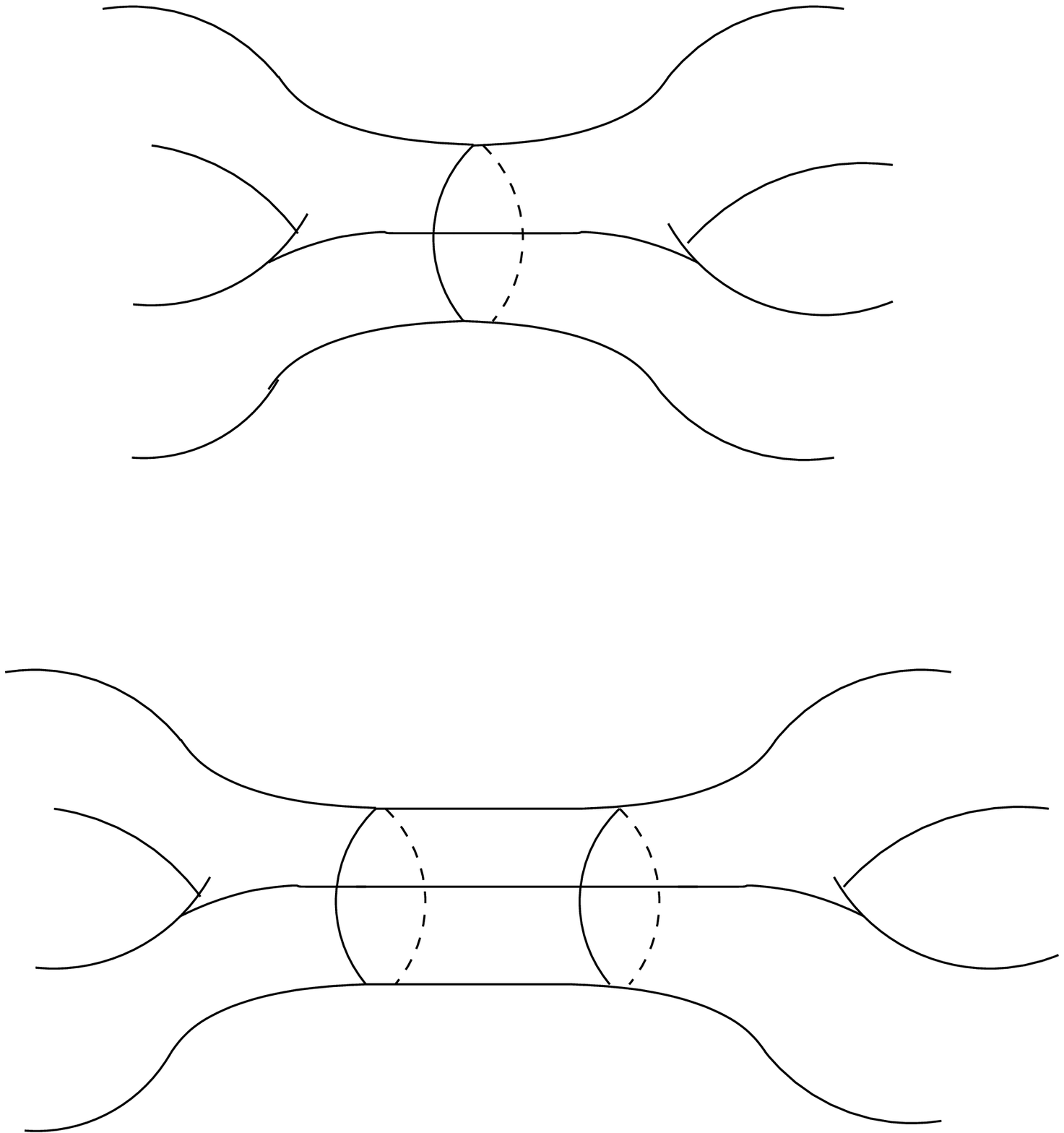}
b)\includegraphics[scale=0.3]{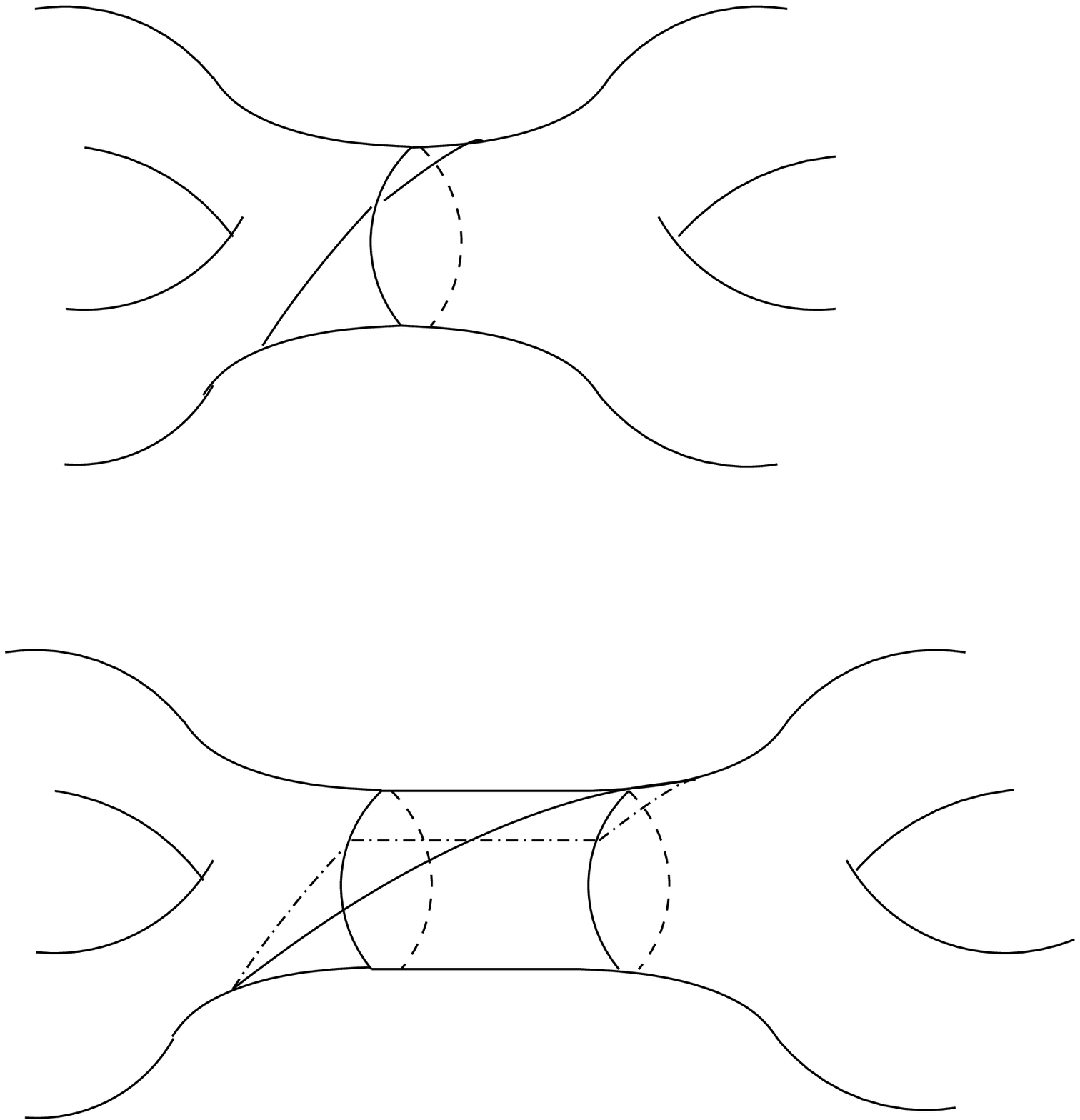}
\caption{\small{Deflection of geodesics through grafting. Case a) corresponds to the situation where $\nu_\lambda=0$, while $\nu_\lambda\neq 0$ in b).}}
\label{geoddefl}
\end{figure}

To investigate the geometrical interpretation of this condition,  one considers an evolving  spacetime obtained by grafting along a single geodesic on the associated Riemann surface $\Sigma_\Gamma=\hyp/\Gamma$ and an observer whose wordline starts at $t=0$ at the initial singularity. It is shown in \cite{icht} that the parameter $\nu_\lambda$ vanishes in this situation if and only if the geodesic associated with $\lambda\in\pi_1(M)$ either does not cross the grafting geodesic or crosses it orthogonally as shown in Figure \ref{geoddefl} a). In this case, the  geodesics associated with $\lambda$ on each constant cosmological time surface  are  not deflected at the grafted strip and their length increases by a constant. The frequency shift of the associated lightray and its directions of emission and return therefore do not depend on the emission time, while the return time depends on it linearly. 

In contrast, if the geodesics on the constant cosmological time surfaces associated with $\lambda\in\pi_1(M)$ cross the grafting geodesic non-orthogonally as shown in Figure \ref{geoddefl} b), they  are deflected at the grafting strip. As the width of the strip remains constant, while the rest of the surface is rescaled with the cosmological time, this deflection depends on the cosmological time $T$  and vanishes for $T\rightarrow\infty$. Their length increase through grafting thus depends non-linearly on the cosmological time and hence the emission time. Consequently, 
 the directions of emission and return of the lightray and its frequency shift depend on the emission time, and the return time becomes a non-linear function of the emission time. 

\section{Reconstructing spacetime geometry from measurements}
\label{sptimegeom}

As demonstrated in the last section, the measurements associated with returning lightrays reflect the geometry of the underlying spacetime and allow the observer to determine some of its geometrical properties. We will now show that they
 allow the observer to reconstruct the {\em full} geometry of the spacetime in  {\em finite} eigentime. 
For this we recall that a spacetime is determined uniquely by the holonomies $h: \pi_1(M)\rightarrow P_3$ modulo simultaneous conjugation with $P_3$. Reconstructing the geometry of the spacetime  is thus equivalent to determining the holonomies for a set of generators $\{\lambda_1,..,\lambda_n\}$ of $\pi_1(M)$ up to simultaneous conjugation with $P_3$.

\subsection{Conformally static spacetime}
We start by considering the case of  conformally static spacetimes and observers whose worldlines start at the initial singularity $M_0$. In this case, the translational components of the holonomies and the parameters $\sigma_\lambda$, $\tau_\lambda$, $\nu_\lambda$ in \eqref{pardef} can be set to zero for all $\lambda\in\pi_1(M)$. Reconstructing the holonomies for a set of generators of $\pi_1(M)$ therefore amounts to determining a set of generators of the  Fuchsian group $\Gamma$ defined by the Lorentzian components of the holonomies.
 A way to achieve this is to reconstruct the Dirichlet region of $\Gamma$. 
 
\begin{figure}[h]
\centering
\includegraphics[scale=0.35]{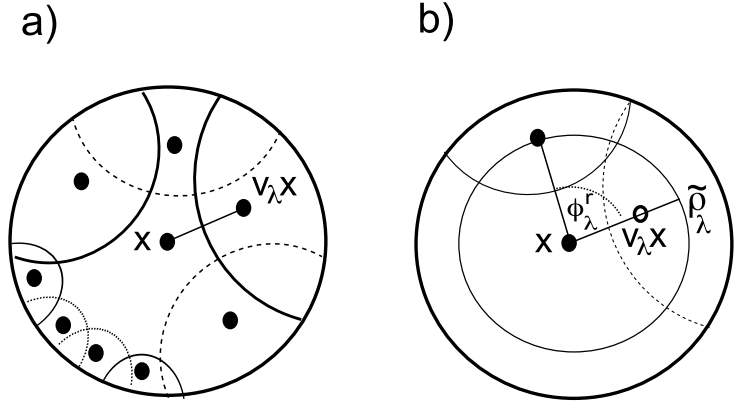}
\caption{\small{ a) Dirichlet region for a Fuchsian group of genus $g=2$.\newline
b) Translated images for evolving spacetimes with deviation angle $\phi_\lambda^r$, modified distance $\tilde \rho_\lambda$.}}
\label{dirichlet}
\end{figure}

 This is the set of points in $\hyp$ whose geodesic distance from a given point $\bx\in \hyp$ is less or equal than the geodesic distance from all its images \cite{katok}
\begin{align}
D_\Gamma(\bx)=\{\by\in\hyp\;|\; d_\hyp(\by,\bx)\leq d_\hyp(\by, v\bx)\;\forall v\in\Gamma\}.\end{align}
It is obtained by constructing the perpendicular bisectors of the geodesic segments $[\bx,v\bx]$ for  $v\in\Gamma$, and intersecting the associated half-planes as shown in Figure \ref{dirichlet} a). The Dirichlet region of a cocompact Fuchsian group $\Gamma$ of genus $g\geq 2$ is a geodesic arc $2k$-gon with  $k\geq 2g$, whose sides are identified pairwise by a set of generators of $\Gamma$. Reconstructing a set of generators of a Fuchsian group $\Gamma$ is thus equivalent to reconstructing a Dirichlet region of $\Gamma$ together with the information about the identification  of its sides\footnote{In the generic case, the sides of the Dirichlet region $D_\Gamma(\bx)$ have different lengths, which allows one to determine directly, which of them are identified by the action of $\Gamma$. However, this is not the case if the dirichlet region has a high degree of symmetry. 
There exist Fuchsian groups which are not isomorphic, have the same Dirichlet region and differ only in the way in which the sides of the Dirichlet region are identified.  In this case, additional information about the identification of sides is needed in order to determine a set of generators of a Fuchsian group $\Gamma$ from its Dirichlet region. I thank R.~C.~Penner for pointing this out to me.}. 

To see how the observer can determine the Dirichlet region of the Fuchsian group $\Gamma$ and the identification of its sides in finite eigentime, we consider the following procedure:
\begin{enumerate}
\item The observer emits light in all directions at a given eigentime $t$. The returning lightrays associated with elements $\lambda\in \pi_1(M)$ return to the observer one by one at different eigentimes $t_\lambda=t+\Delta t_\lambda$. For each returning lightray, the observer measures the eigentime $\Delta t_\lambda$ elapsed since the emission  and the direction from which the light returns. The former allows him to determine the geodesic distance $\rho_\lambda=d_\hyp(\bx,v_\lambda\bx)$ between his velocity vector $\bx\in\hyp$ and its image $v_\lambda\bx\in\hyp$ via \eqref{statvalue}. The latter corresponds to the tangent vector at $v_\lambda\bx$ to the geodesic segment $[\bx, v_\lambda\bx]$ in $\hyp$.  Hence, given the observer's velocity vector $\bx$, the measurement of the return time and direction of return allow him to determine the image $v_\lambda\bx$.

\item For each returning lightray, the observer can thus construct the perpendicular bisector of the geodesic segment  
$[\bx,v_\lambda\bx]$ in $\hyp$ as shown in Figure \ref{dirichlet} a). 
 After a finite number of returning lightrays, the perpendicular bisectors of the associated geodesic segments $[\bx, v_\lambda \bx]$ close to form a geodesic arc polygon $P\subset \hyp$ as shown in Figure \ref{dirichlet} a). If $r=\text{max}\{d_{\hyp}(\bx,\bp)\;|\; \bp\;\text{corner of}\; P\}$, the perpendicular bisectors of images $v_\lambda\bx$ with $d_\hyp(\bx, v_\lambda\bx)>2r$ cannot intersect $P$ and therefore do not affect the Dirichlet region. This implies that lightrays returning after a time $\Delta t=t(e^{2r}-1)$ are irrelevant, and the observer can reconstruct the Dirichlet region $D_\Gamma(\bx)$ in finite eigentime $\Delta t=t(e^{2r}-1)$ .

\item After determining the Dirichlet region, the observer sends out a finite number of additional lightrays into the directions associated with the geodesic arcs that form the boundary of the Dirichlet region and records from which directions the associated lightrays return. This allows him to conclude which sides of the Dirichlet region are identified by $\Gamma$. Together with the Dirichlet region, this information allows him to reconstruct a set of generators of  $\Gamma$\footnote{I thank R.~C.~Penner and several  other participants of the workshop ``Chern-Simons Gauge Theory: 20 years after" for discussions of this issue.}. \end{enumerate}

This  procedure  allows the observer  to determine a set of generators of the Fuchsian group $\Gamma$ and hence to reconstruct the full geometry of the spacetime in finite eigentime. Moreover, it is shown in  \cite{icht} that the observer does not need to know his velocity vector $\bx$ to do so, as a change of the vector $\bx\mapsto v\bx$ with $v\in SO^+(2,1)$ amounts to simultaneous conjugation of all elements of  $\Gamma$ with $v$.

\subsection{Evolving spacetimes}

We now consider the case of a general observer in an evolving spacetime. In this case, the observer  can proceed as if the spacetime was conformally static and measure the return time and direction for each returning lightray as outlined in the last subsection. 
The return direction then depends on the emission time and is modified with respect to the return direction of the associated conformally static spacetime by the angle
\begin{align}\label{phimod}
\phi^r_\lambda(t)=\arctan\left(\nu_\lambda/(t+\sigma_\lambda)\right)=\nu_\lambda/t+O(1/t^2).
\end{align}
Moreover, the observer can use expression \eqref{statvalue} for the return time in a conformally static spacetime to assign to each returning lightray a modified distance $\tilde \rho_\lambda(t)$ defined by
\begin{align}\label{evoldef}
\Delta t(t,\bx,\bx_0, h(\lambda))=t(e^{\tilde \rho_\lambda(t)}-1).
\end{align} 
In terms of the hyperbolic distance variable $\rho_\lambda=d_{\hyp}(\bx, v_\lambda\bx)$ of the associated static spacetime  this modified distance is given by
\begin{align}\label{rhomod}
\tilde \rho_\lambda(t)=&\ln\left(e^{\rho_\lambda} -\tau_\lambda / t+\sinh\rho_\lambda\left( \sqrt{(1+\sigma_\lambda/t)^2+\nu_\lambda^2/t^2}-1\right)\right)\nonumber\\
=&\rho_\lambda+({\sigma_\lambda(e^{\rho_\lambda}-1)-\tau_\lambda})/{t}+O(1/t^2).
\end{align}
The  observer can now act as if the spacetime was conformally static and construct a geodesic arc polygon as outlined in the last subsection, using the variables $\tilde \rho_\lambda(t)$ and the return directions obtained from his measurements.  The images of the velocity vector $\bx\in \hyp$
obtained this way and, consequently, the resulting perpendicular bisectors will be translated with respect to the conformally static case  as indicated in Figure \ref{dirichlet} b). The observer thus constructs a
deformed geodesic arc polygon $\tilde P(t)$ which approaches the Dirichlet region of the associated conformally static spacetime in the limit $t\rightarrow\infty$. 
 
To determine the holonomies along a set of generators of the fundamental group, the observer can now repeat the measuring procedure outlined in the last subsection several times and obtains a sequence of deformed polygons $\tilde P(t_1), \tilde P(t_2),..., \tilde P(t_n)$ in $\hyp$. By observing the change of the polygons with the emission time, the observer can extrapolate to the limit $t\rightarrow \infty$ to recover the Dirichlet region of the associated conformally static spacetime and, after a finite number of additional measurements, the identification of its sides.  
Using formulas \eqref{phimod} to  \eqref{rhomod}, he can then determine the associated parameters $\sigma_\lambda,\tau_\lambda,\nu_\lambda$. Via \eqref{pardef} he then obtains  the holonomies $h(\lambda)$  for a set of generators of the fundamental group $\pi_1(M)$ and hence the full geometry of the spacetime in finite eigentime.

\section{Concluding remarks}
\label{outlook}

We showed how the description of flat MGH 3d Lorentzian manifolds in terms of  their universal cover can be used to obtain interesting physics in (2+1)-dimensional gravity.  By considering an observer who probes the geometry of the spacetime by emitting returning lightrays, we defined several  measurements that could be made by such an observer: the eigentime elapsed between the emission of the lightray and its return, the directions into which light is emitted and from which it returns as well as the frequency shift between the emitted and returning lightray.

We gave explicit expressions for these measurements in terms of the holonomy variables which parametrise the flat Lorentzian 3d manifolds  arising in (2+1)-gravity
 and  play a central role in the quantisation of the theory.  Moreover,  we demonstrated how an observer can use these measurements to determine the holonomy variables and thus reconstruct the full geometry of the spacetime in finite eigentime.  

The results serve a concrete and non-trivial example
 in which concrete physics questions and conceptual issues of (quantum)  gravity can be investigated  \cite{icht}.  It would also be interesting to generalise them  to more realistic scenarios   relevant to cosmology such as 
 observations of external light sources or background radiation emitted near the initial ``big bang" singularity.  Finally, one of the main motivations of this work is its application to
a quantum theory of (2+1)-gravity. This would offer the prospect of  investigating  realistic 
 physical measurements  in a fully and rigourously quantised theory of gravity.

\section*{Acknowledgements}

I thank the organisers of the workshop ``Chern-Simons Gauge Theory: 20 years after" in which this work was presented, and I am grateful to its participants for helpful comments and suggestions.  In particular, I thank  R.~C.~Penner for helpful remarks regarding the measurement of the Dirichlet region. This work was supported by the Emmy Noether fellowship ME 3425/1-1 of the German Research Foundation (DFG).

\end{document}